\documentclass[conference]{IEEEtran}
\IEEEoverridecommandlockouts
\usepackage{cite}
\usepackage{amsmath,amssymb,amsfonts}
\usepackage{enumitem}
\usepackage{algorithmic}
\usepackage[ruled,vlined]{algorithm2e}
\usepackage{graphicx}
\usepackage{textcomp}
\usepackage{xcolor}
\usepackage[numbers]{natbib}
\def\BibTeX{{\rm B\kern-.05em{\sc i\kern-.025em b}\kern-.08em
    T\kern-.1667em\lower.7ex\hbox{E}\kern-.125emX}}

\begin{document}



\title{Generative Intelligence Systems in the \\ Flow of Group Emotions}


\author{
\IEEEauthorblockN{
Fernando Koch\IEEEauthorrefmark{1}, 
Jessica Nahulan\IEEEauthorrefmark{2}, 
Jeremy Fox\IEEEauthorrefmark{2}, 
Martin Keen\IEEEauthorrefmark{2}}
\IEEEauthorblockA{\IEEEauthorrefmark{1}Florida Atlantic University, USA\\
kochf@fau.edu}
\IEEEauthorblockA{\IEEEauthorrefmark{2}IBM, Canada / USA\\
jessican@ibm.com, \{jeremyf, mkeen\}@us.ibm.com}
}

\maketitle


\begin{abstract}


Emotional cues frequently arise and shape group dynamics in interactive settings where multiple humans and artificial agents communicate through shared digital channels. While artificial agents lack intrinsic emotional states, they can simulate affective behavior using synthetic modalities such as text or speech. This work introduces a model for orchestrating emotion contagion, enabling agents to detect emotional signals, infer group mood patterns, and generate targeted emotional responses. The system captures human emotional exchanges and uses this insight to produce adaptive, generative responses that influence group affect in real time. The model supports applications in collaborative, educational, and social environments by shifting affective computing from individual-level reactions to coordinated, group-level emotion modulation. We present the system’s architecture and provide experimental results that illustrate its effectiveness in sensing and steering group mood dynamics.

\end{abstract}


\section{Introduction}
\label{sec:introduction}


Computational Intelligence Systems have become adept at interpreting and responding to human inputs due to recent advances in natural language processing, sentiment analysis, and machine learning \cite{broekens2023fine}. These capabilities allow artificial agents to recognize users’ affective cues and generate emotionally appropriate interactions \cite{hohenstein2020ai}. However, most existing systems remain narrowly focused, particularly on dyadic exchanges and reactive emotion modeling and reactive emotion modeling, which limits their ability to support dynamic, group-level emotional reasoning \cite{bosse2015agent, zhang2023review}.  In particular, a phenomenon known as \textit{emotional contagion}, or the negative propagation of emotion in online settings, is a growing concern due to its disruptive potential \cite{vsuvakov2012agent, prollochs2021emotions}. This underscores the need for approaches that can regulate shared emotional dynamics in group settings. This research is driven by the question:

\begin{quote}
\textit{How can intelligent agents orchestrate adaptive and generative emotional responses to promote positive affective convergence and mitigate negative emotion contagion in multi-human to multi-agent communication environments?}
\end{quote}

We introduce a generative AI (genAI)-based orchestration system designed to influence group emotional dynamics in multi-party interactions. The system continuously senses distributed emotional signals, detects prevalent moods across the group, and orchestrates adaptive agent responses that promote emotional alignment. The core innovation lies in leveraging decentralized mood-sensing and genAI pipelines to enable collective emotional reasoning among agents, allowing agents to align their affective responses and collaboratively influence the group’s emotional trajectory.

This model extends the emotional depth of digital interactions, allowing agents to function not only as individual assistants but as collaborative social actors in emotionally rich environments. Agents can help maintain healthier group affect \cite{barsade2015group, naveenan2018impact, delice2019advancing} by actively fostering positive emotional alignment (affective convergence) and damping the spread of negative sentiment. This approach aims to shift the design of intelligent agents from reactive empathy to proactive, coordinated emotional orchestration in multi-agent, multi-user systems.

This work contributes to the state of the art by providing:

\begin{itemize}
    \item \textbf{A reference architecture} for the generative orchestration of emotion contagion in multi-human, multi-agent systems, supporting distributed mood sensing, emotional pattern grouping, and coordinated agent response generation.

    \item \textbf{A method for real-time emotional alignment} that enables conversational agents to reason collectively about group mood and adapt their responses using generative AI pipelines conditioned on both local and global affective contexts.

    \item \textbf{An approach to configuration-aware emotional modulation}, demonstrating how the system can operate effectively across various interaction topologies—ranging from dyadic support to coordinated agent collectives in social networks.
\end{itemize}

\section{Background}
\label{sec:background}


A seminal study on emotional contagion by Barsade (2002)\cite{barsade2002ripple} demonstrated that group members tend to ``catch each other’s moods'', with positive contagion improving collaboration and task performance, while negative moods amplify conflict. In particular, unpleasant or negative emotions often propagate more strongly and quickly than positive ones, creating a ``ripple effect'' that can undermine the cohesion of the group. In online environments, this phenomenon persists on scale: a large Facebook experiment showed that reducing positive content in users’ news feeds led those users to post more negative updates (and vice versa), providing experimental evidence of mass emotional contagion through digital communication \cite{kramer2014experimental}. This contagion can erode trust and participation in multi-user interactions, that is, a sarcastic or angry remark by one user can trigger an ``epidemic of grumpiness'' that dampens the experience of the entire group.

Current conversational AI systems, however, are ill-equipped to manage these group-level affective dynamics~\cite{picard2000affective, hatfield2011emotional}. Most virtual agents and chatbots operate in dyadic (one-on-one) settings and react to the emotions of a single user in a strictly individualistic way \cite{jiang2023communitybots, amiot2025chatbots}. As a result, they lack awareness of the broader mood in a multi-user conversation and cannot prevent negative emotions from cascading across participants. Researchers have noted that the deployment of chatbots as active participants in multiparty dialogues remains largely underexplored, requiring new capabilities such as managing turn-taking, roles and social context among multiple people \cite{dohsaka2014effects}.

Until recently, emotionally intelligent agents for health and education were almost exclusively designed for one-on-one interactions \cite{nordberg2019designing}. This narrow focus is a critical limitation in settings (e.g. group chats, online forums, multiplayer games, collaborative teams) where emotion contagion can quickly spiral and derail the collective experience \cite{barsade2002ripple, kramer2014experimental}. In summary, the challenge is clear: negative affect can amplify and spread in group interactions, but today’s AI lacks the group-level emotional reasoning needed to detect and mitigate these contagious downturns before they undermine user trust and participation.

Bosse et al. (2015)~\cite{bosse2015agent} introduced an agent-based model of emotion contagion within teams, integrated into an ambient intelligence system that monitors group sentiment and suggests supportive interventions to preempt ``downward emotion spirals'' . This work formalized how an automated team assistant could track the emotional levels of multiple humans and issue group-level support actions (e.g., encouraging messages) when collective morale dips. Subsequent simulations expanded on these ideas, modelling the dynamics of emotion spread in agent networks, showing how mood contagion could be predicted and controlled in principle. 

In practice, a few domain-specific systems have demonstrated the benefits of coordinated, emotionally aware agents in multi-user settings. In educational technology, socially intelligent tutors have been augmented with positive socio-emotional strategies to foster a constructive group climate. For example, Dohsaka et al. (2014)~\cite{dohsaka2014effects} built a multiparty quiz game system involving two humans and two agent facilitators. They found that the presence of participants from empathic agents significantly increased user satisfaction and even the number of user contributions to the conversation. These studies illustrate that a carefully orchestrated agent response can promote engagement and positive emotional convergence in a group.

More recently, researchers have started exploring multi-agent setups explicitly for group emotional support. Nordberg et al. (2019)~\cite{nordberg2019designing} presented Terabot, a chatbot designed to guide peer support conversations among users with ADHD in an online program. Acting as a facilitator, the bot provided structure and encouragement in the group chat, which the participants found helpful to keep the discussions on track. This multi-chatbot counseling system led to higher user engagement and notable linguistic convergence compared to a one-on-one chatbot, indicating a stronger rapport and a sense of support. However, this solution used agents with predefined \emph{facilitator} and \emph{peer} roles, meaning the bots followed a scripted division of labor with minimal real-time coordination or ``collective reasoning'' among themselves. Even considering this limitation, participants in this experiment felt greater social support and motivation to cope, demonstrating how multiple agents working together can shape a more positive and resilient group mood.

While these efforts are promising, there remains a significant gap in achieving truly adaptive, generative emotional orchestration for multi-human/multi-agent environments. Most existing systems rely on scripted rules or fixed role strategies to handle group emotions. For example, some mental health chatbots detect when a user expresses negative feelings and then trigger a prewritten empathetic prompt or suggest contacting a crisis hotline. Such rule-based interventions can be helpful, but lack the flexibility and creativity that modern generative AI can provide.

Currently, there is no publicly documented platform that supports a decentralized network of agents capable of sharing emotional signals and coordinating group mood responses in real time. The focus in AI-driven emotional support has been on empathy for a single agent or on centralized settings where an intelligent agent attempts to facilitate or monitor an entire group \cite{tavanapour2020conversational}. This leaves a technological gap in enabling multiple agents to behave as a team of socially savvy actors, dynamically harmonizing their responses to manage group affect in an open-ended way.

Our system also aligns with architectural strategies from safety-critical domains. For instance, prior work on augmented reality recommendation systems in emergency scenarios~\cite{beloglazov2017augmented} demonstrated the value of context-aware, distributed sensor input to dynamically guide user actions in real time. This approach parallels our use of affective signals for dynamic, emotion-driven agent coordination, highlighting a shared need for responsiveness, decentralization, and adaptive orchestration across critical and emotionally intensive settings.

\section{Proposal}
\label{sec:proposal}


We propose a generative orchestration system that enables a network of conversational agents to collaboratively regulate group emotion in real time. The innovation lies in combining decentralized mood sensing with generative emotional response coordination, allowing agents to act not as isolated responders but as an affectively aware collective capable of influencing group mood dynamics.

Our framework allows each agent to perceive emotional signals from the humans with whom it engages and to share a summary of its local ``mood'' readings with other agents~\cite{asghar2018affective}. This distributed emotional awareness supports collective emotional reasoning, in which agents form a shared representation of the affective state of the group through real-time consensus.

Based on this shared model, agents generate adaptive responses using generative AI pipelines conditioned on both the local interaction context and global emotional goals~\cite{wang2025rlver}. For example, agents can introduce encouraging dialogue, express empathy, or inject subtle humor in a coordinated fashion when negativity is detected, guiding participants toward emotional convergence without explicit instruction.

\begin{figure}[h!]
    \centering
    \includegraphics[width=0.85\linewidth]{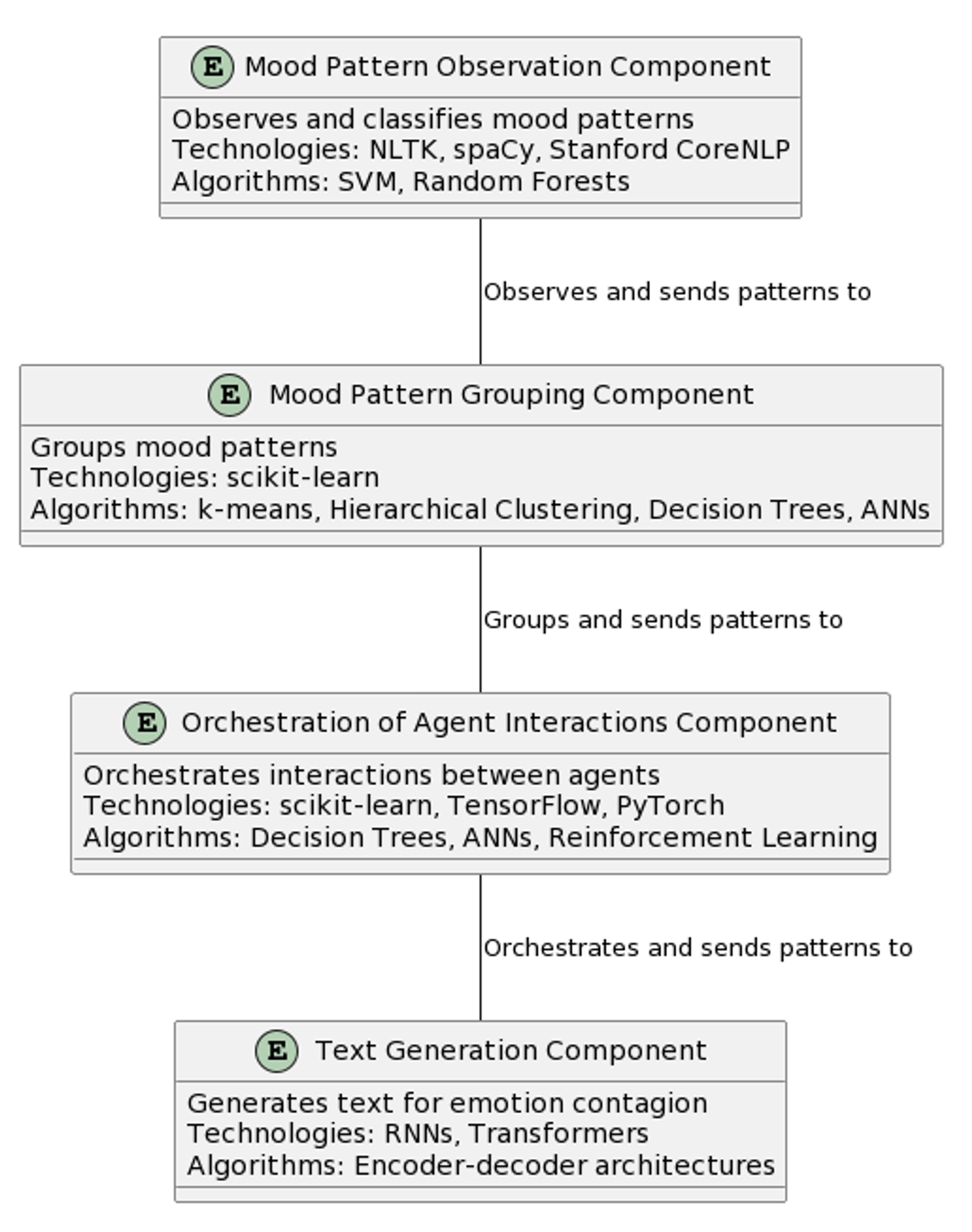}
    \caption{Solution Overview}
    \label{fig:sol-overview}
\end{figure}

The architecture is described in Fig.~\ref{fig:sol-overview} and works as follows. Each conversational agent independently observes emotional cues from human interactions through the \emph{Mood Pattern Observation Component}, which classifies user mood using NLP pipelines and supervised learning algorithms. These local mood patterns are passed to the \emph{Mood Pattern Grouping Component}, which aggregates and clusters emotional signals from agents using unsupervised processing methods. This allows the system to identify dominant affective trends and emotional divergence in the group.

\begin{figure}[h!]
    \centering
    \includegraphics[width=0.85\linewidth]{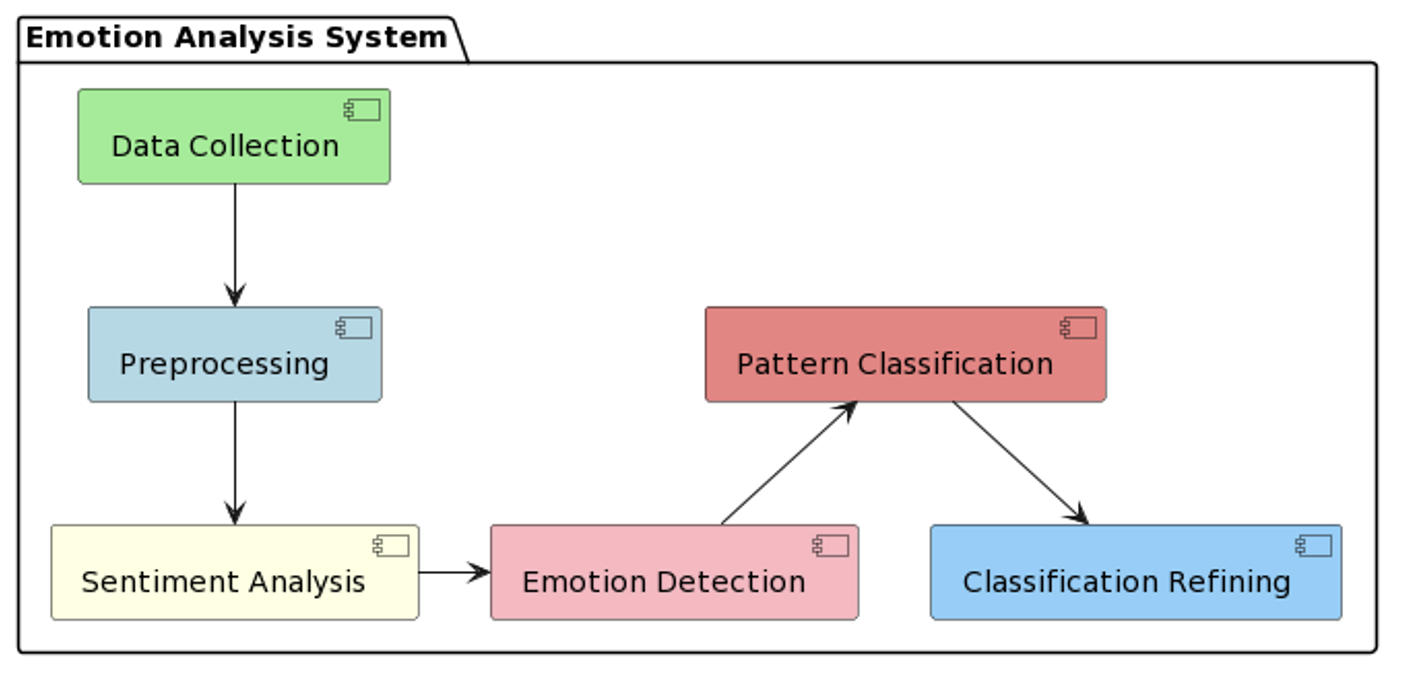}
    \caption{Mood Pattern Observation Module}
    \label{fig:mood-pattern-obs}
\end{figure}

Fig.~\ref{fig:mood-pattern-obs} depicts the architecture of \emph{Mood Pattern Observation Component}, which operates through the following sequential steps:

\begin{itemize}
    \item \emph{Data Collection and Preprocessing}: collects human-agent conversations and executes he methods to clean, normalize, and prepare for downstream analysis.

    \item \emph{Sentiment Analysis}: analysis each utterance to extract valence and polarity scores, providing a basic emotional tone for the text.

    \item \emph{Emotion Detection}: applies emotion lexicons and classifiers to identify specific emotions (e.g., joy, anger, sadness) expressed in the conversation.

    \item \emph{Pattern Classification}:  group detected emotional states into \textit{mood patterns} based on their intensity and temporal continuity, capturing the emotional flow over time.

    \item \emph{Classification Refinement}: apply context-aware rules, eventually combined with GenAI-powered analysis, to refine the predictions, improve accuracy, and robustness.
\end{itemize}

This component enables each agent to maintain a fine-grained, real-time understanding of the emotional state of its human counterpart while also sharing local mood observations with other agents in the system. The inter-agent sharing process provides the foundation of collective emotional reasoning, allowing agents to combine their individual insights into a broader, group-level emotional model.

\begin{figure}[h!]
    \centering
    \includegraphics[width=0.85\linewidth]{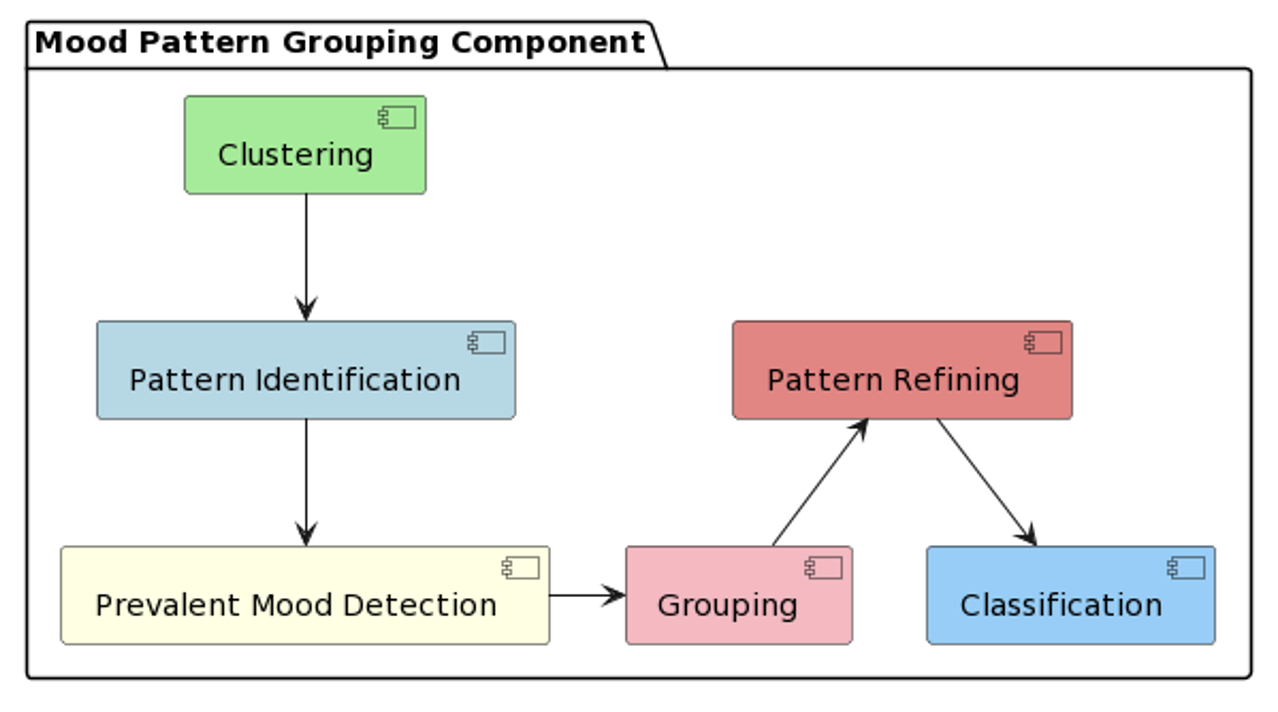}
    \caption{Mood Pattern Grouping Module}
    \label{fig:mood-pattern-grouping}
\end{figure}

Then, the \emph{Mood Pattern Grouping Component} (see Fig.~\ref{fig:mood-pattern-grouping}) executes the following steps:

\begin{itemize}
    \item It begins by clustering observed mood patterns based on similarity using unsupervised learning algorithms such as k-means or hierarchical clustering.
    
    \item The resulting clusters are passed to the \emph{Pattern Identification} module, which detects common emotional trajectories across conversations.
    
    \item The \emph{Prevalent Mood Detection} module identifies dominant emotional trends based on the frequency, intensity, and duration of the clustered mood data.
    
    \item The \emph{Grouping Module} organizes conversations according to their shared emotional tone.
    
    \item These initial groupings are refined by the \emph{Pattern Refining} module using machine learning models such as decision trees or artificial neural networks to enhance granularity and cohesion.
    
    \item Finally, the \emph{Classification} module assigns each conversation group to predefined affective categories—such as positive, negative, or neutral—based on the refined features.

\end{itemize}

This component enables the system to transition from local emotional sensing to shared emotional awareness. It supports collective reasoning by providing each agent with a contextual map of emotional convergence and divergence across the group. This shared understanding allows agents to act in coordination, targeting emotionally divergent users with responses that support affective alignment.

\begin{figure}[h!]
    \centering
    \includegraphics[width=0.85\linewidth]{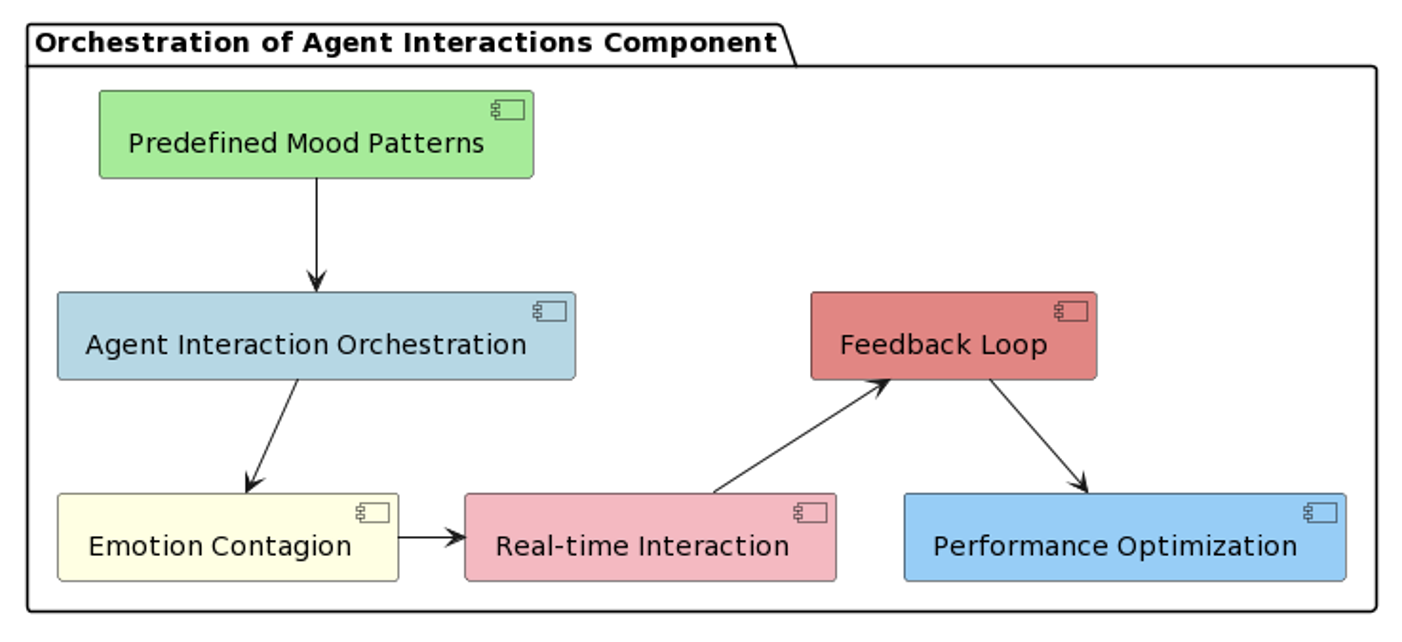}
    \caption{Orchestration of Agent Interactions Component}
    \label{fig:orch-agent-interaction}
\end{figure}

Fig.~\ref{fig:orch-agent-interaction} depicts the \emph{Orchestration of Agent Interactions Component}, which coordinates how agents modulate emotional responses in alignment with group mood dynamics. This component  operates through the following steps:

\begin{itemize}
    \item Given a set of \emph{Predefined Mood Patterns}, which represent common affective states such as happiness, sadness, anger, or fear. These patterns are derived from historical mood observations and inform agent strategies.
    
    \item The \emph{Agent Interaction Orchestration} module combines these mood patterns with the current group affective state to determine how agents should respond, applying decision trees to select coordinated agent behaviors during ongoing conversations.
    
    \item The selected strategy is implemented through the \emph{Emotion Contagion Component}, where agents adapt their conversational tone, content, and style using generative models to nudge user mood toward emotionally aligned states.
    
    \item There is a continuous \emph{Feedback Loop} evaluates the emotional outcomes of agent interactions.
    
    \item Based on these observations, the \emph{Performance Optimization} module refines orchestration strategies using reinforcement learning techniques, and  adjusts parameters like message timing, language style, or agent allocation through adaptive tuning.
\end{itemize}

This component shifts emotional reasoning from static, pre-scripted planning to dynamic, responsive coordination. It empowers agents to operate as an emotionally synchronized collective, maintaining affective coherence across interactions while adapting in real time to evolving user sentiment.

\subsection*{Putting Things Together}

The entire system functions as a coordinated emotional reasoning pipeline that enables conversational agents to understand, share, and influence collective affect in real time. What begins as individual emotional perception becomes, through successive stages of aggregation and orchestration, a structured, system-wide capacity for emotional modulation.

This architecture addresses a critical limitation in current AI systems: the inability to transition from isolated, reactive emotion recognition to proactive, collaborative emotion shaping. Traditional approaches have treated emotional responses as static or individual phenomena. In contrast, this model enables agents to reason about mood at the group level, share their affective perceptions, and act in unison to steer the collective emotional trajectory~\cite{mao2024multi}.

The orchestration framework embodies the idea of collective intelligence applied to emotion. Agents contribute to an emergent, coordinated strategy that supports emotional convergence and mitigates negative contagion. This architecture introduces a foundation for a new class of emotionally-aware, socially intelligent agent systems equipped for  domains where emotional alignment and group cohesion are integral to sustained engagement.

\section{Configurations}
\label{sec:configs}


We consider different deployment configurations in which the proposed generative orchestration system may operate.

\subsection{Dialytic Interactions (H–M–Ag)}

\begin{figure}[h!]
    \centering
    \includegraphics[width=0.95\linewidth]{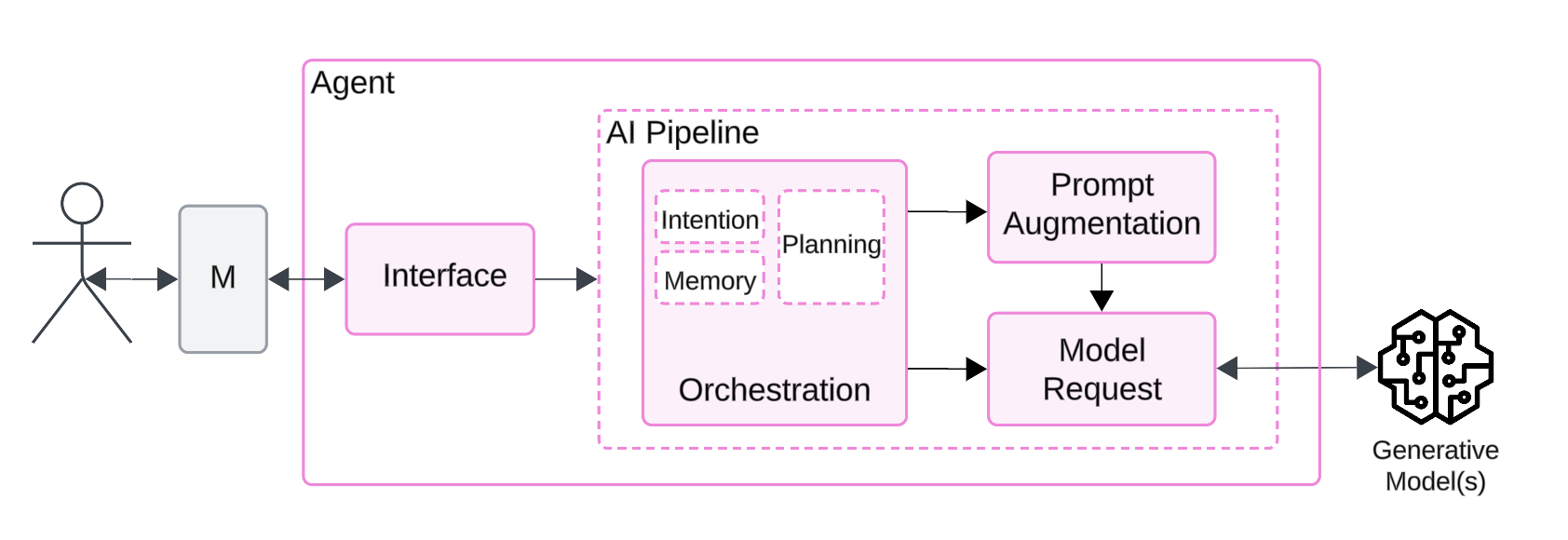}
    \caption{Configuration with a basic dyadic interaction model}
    \label{fig:config-hma}
\end{figure}

At the basic level, there is the dyadic interaction loop Human(H)–Medium(M)–Agent(Ag) (see Fig.~\ref{fig:config-hma}). In this configuration, the human user interacts with a conversational agent through a communication medium. An agent can be considered to be composed of an interface connected to an internal AI pipeline comprising orchestration, intention, memory,  planning modules, and a pipeline for prompt augmentation and model invocation. These agents generate responses by orchestrating calls to the underlying generative models to complement and contextualize information. However, the specific composition of the agent architecture is not central to our discussion.

In dyadic configurations, emotional contagion is typically constrained. The agent's affective behavior is shaped exclusively by the mood of the individual user, and there is no direct exposure to emotional signals from other users. As a result, emotional alignment occurs in isolation, lacking the benefit of collective affective awareness or group-level feedback. In this environment, our system introduces capabilities that expand the emotional intelligence of the agent beyond the local dyad. In particular, \emph{Mood Pattern Observation Component} continues to detect and classify emotional cues from the user in real time. These local emotional signals are then shared across the agent network and processed by the \emph{Mood Pattern Grouping Component}, which identifies emergent mood trends across different users and contexts.

This cross-agent exchange allows even isolated agents to act with knowledge of prevalent emotional climates. When an agent detects negative emotion from its user, it can adjust its generative responses not only based on local history but also in alignment with affective strategies used successfully in other similar contexts. The \emph{Orchestration of Agent Interactions Component} adapts responses using strategies such as emotional mirroring, gentle humor, or reframing to foster emotional convergence with broader user trends.

Therefore, the system transforms even conventional one-to-one agent configurations into emotionally coherent nodes within a distributed, affect-aware network, enabling localized experiences to benefit from collective intelligence.

\subsection{Social Network with Personal Assistance (Ag-H–M)}

\begin{figure}[h!]
    \centering
    \includegraphics[width=0.95\linewidth]{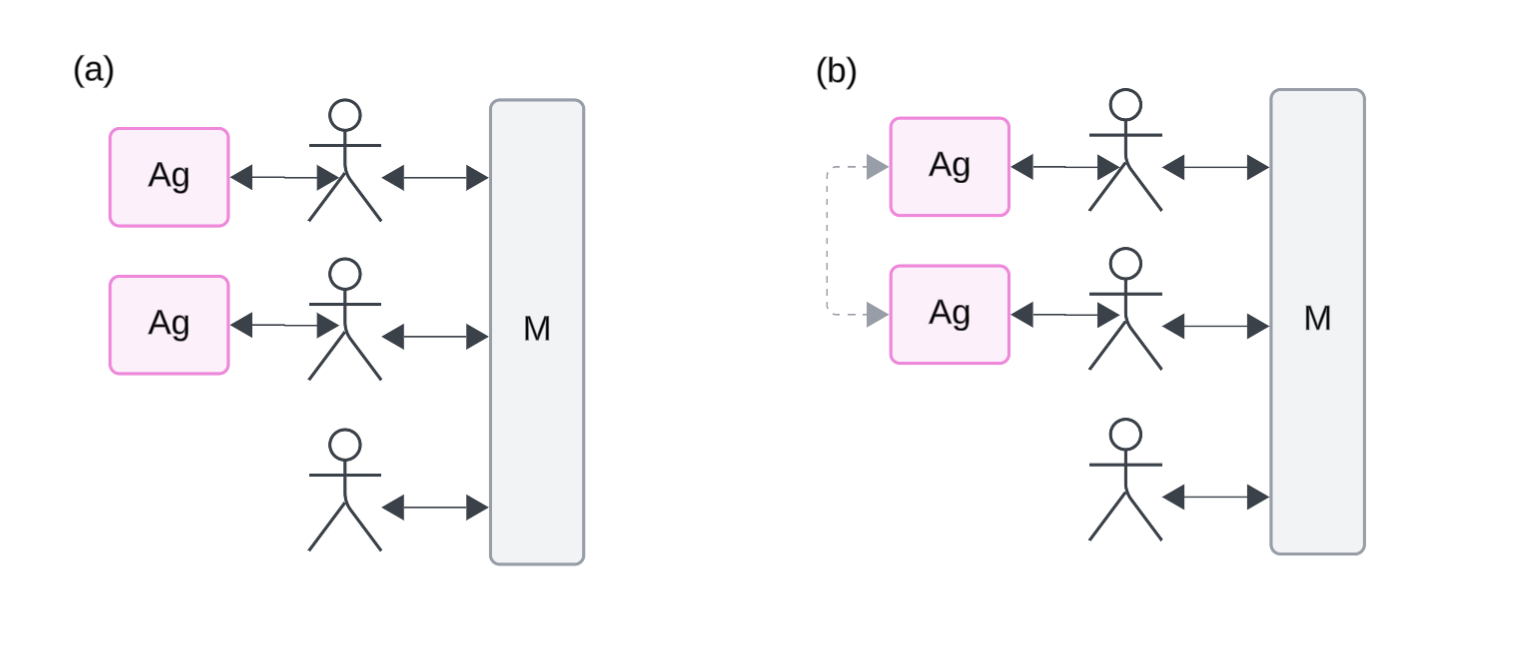}
    \caption{Configurations where agents act as personal assistants without direct access to the medium. (a) No agent inter-communication. (b) Agents communicate in the background.}
    \label{fig:config-ahm}
\end{figure}

In configurations Agent(Ag)-Human(H)–Medium(M) (see Fig.~\ref{fig:config-ahm}), conversational agents serve as personal perceptual assistants for the individual users, but do not interact directly with the shared medium. These scenarios occur on digital social platforms and collaborative tools, where the medium of interaction is shared among users, but each user can also engage a background agent for cognitive or emotional support.

In variation Fig.~\ref{fig:config-ahm}(a), each human user interacts with a shared environment (e.g., social network or messaging platform), while their respective agent provides individualized support. However, these agents do not have access to the content that circulates in the medium. Emotional contagion in this setup occurs primarily through human-to-human interactions, while agents can only infer emotional shifts indirectly through their user’s responses. This creates a blind spot in affective awareness and limits the agent's ability to support proactive emotion modulation.

Our system addresses this limitation by enabling agents to pool mood observations through shared emotional models, even if they do not have access to the full content of the medium. Through \emph{Mood Pattern Observation Component}, agents monitor the emotional expressions of their assigned users. These local signals are aggregated via the \emph{Mood Pattern Grouping Component} across agents to infer prevalent affective trends among the broader user base. In this way, even agents with partial information can collectively reason about the mood of the group.

The variation depicted in Fig.~\ref{fig:config-ahm}(b) expands this concept by allowing agents to communicate with each other in the background. Here, emotional contagion becomes a multilayered phenomenon: while humans influence each other through the shared medium, agents also exchange emotional cues and response strategies through the orchestration pipeline. This configuration models emergent social behavior in mediated environments, such as support groups or comment threads.

Our system enhances this interaction by activating the \emph{Orchestration of Agent Interactions Component}, which synchronizes emotional responses among agents based on the mood of the consensus group. This enables agents to deliver coordinated support, recommending uplifting interactions, suggesting de-escalation when tensions rise, or reinforcing affirming sentiments, and thus steering collective emotional trajectories without requiring direct access to the medium.

In both configurations, the proposal shifts the role of agents from isolated affective mirrors to cooperative emotional mediators. It enables emotionally intelligent behavior in contexts where agents act as silent observers and advisers, extending the reach of affective computing into domains where emotional nuance and social harmony are essential but difficult to instrument directly.

\subsection{Humans and Agents Participating in Social Networks ([H|Ag]–M)}
\label{sec:config-hma}

\begin{figure}[h!]
    \centering
    \includegraphics[width=0.95\linewidth]{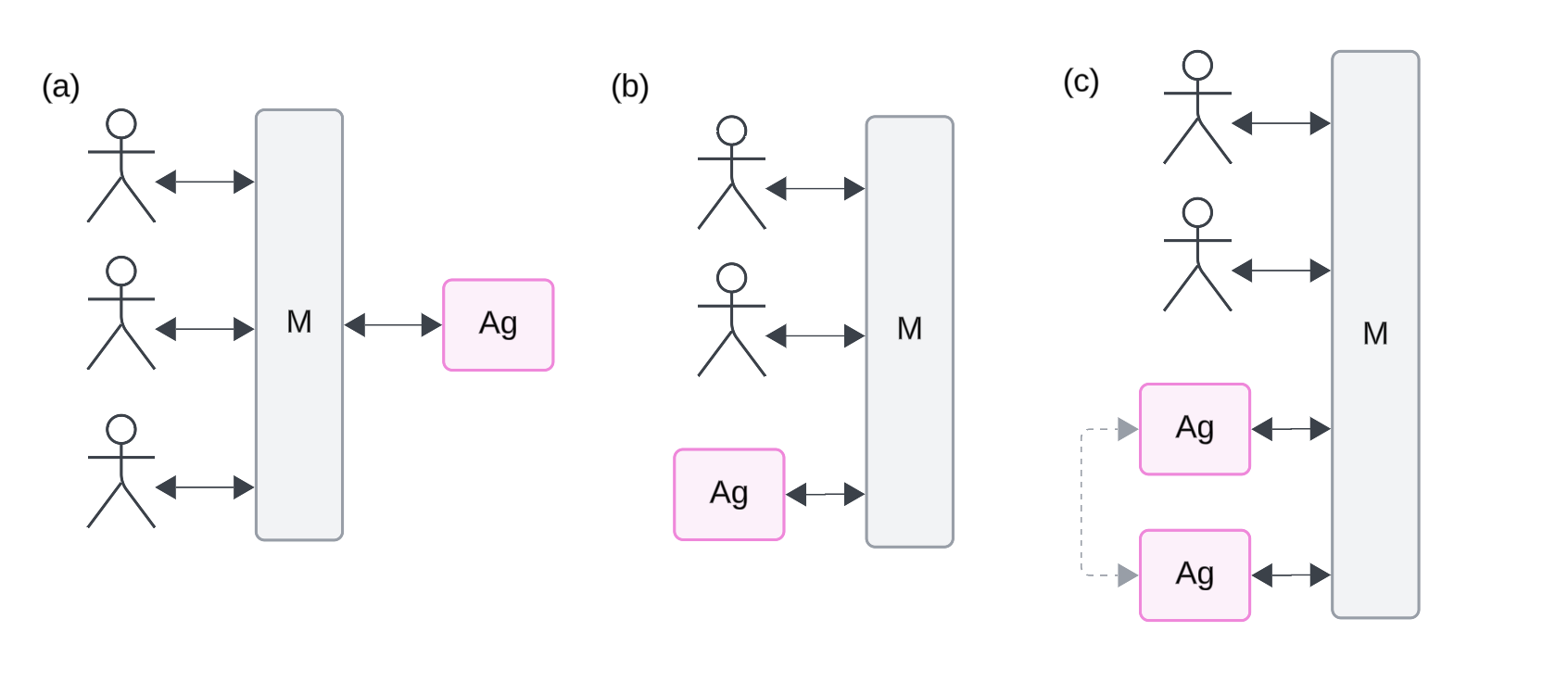}
\caption{
Configurations with Agents Participating in Social Networks.
(a) Agents observe human-human interaction within the shared medium and inject responses to influence mood without direct engagement.
(b) Agents actively participate alongside humans in mediated communication.
(c) Multiple agents interact with users within the shared medium and coordinate among themselves through a hidden backchannel to synchronize strategies.
}
    \label{fig:config-ham}
\end{figure}

Figure~\ref{fig:config-ham} presents the configuration where Humans(H)|Agents(Ag)-Medium(M), representing the situation where both humans and agents exchange messages over a shared platform. In these settings, agents have varying degrees of visibility and engagement, making them well-positioned to influence emotion contagion at scale.

The variation depicted in Figure~\ref{fig:config-hma}(a) involves agents that are not engaged in direct interaction with users, but observe ongoing conversations via the shared medium and can inject messages visible to all participants. This indirect presence enables agents to sense prevailing emotional trends, detect negative affect spirals, and strategically issue mood-regulating prompts into the conversation stream, such as humor, encouragement, or clarification. This design is well suited for forums or comment threads, where agents can act as moderators or tone-steering entities.

Figure~\ref{fig:config-ham}(b) presents a variation where agents participate alongside humans, directly participating in affective exchanges. Each agent contributes to the ongoing dialogue by aligning its tone and behavior to the detected mood state of its conversation partner while remaining contextually aware of the general sentiment of the group. In this setup, the orchestration system enables generative affective responses tailored to local and group mood dynamics, thus enhancing emotional alignment during shared experiences. This is appropriate for live discussion groups or multiplayer environments.

Figure~\ref{fig:config-ham}(c) depicts the configuration where  multiple agents interact with different users in the shared medium while also maintaining a backchannel for inter-agent communication. This backchannel allows agents to synchronize emotional strategies, share observed mood patterns, and agree on joint responses in real time. This structure leads to emergent group-level mood regulation, where agents collectively work to diffuse negative affect and promote affective convergence without requiring direct instruction or user intervention. This configuration is suitable for decentralized emotional reasoning in large-scale social platforms.

These scenarios illustrate how the proposed system adapts to varying agent roles. In each case, the core innovation of the system enables proactive mood shaping in complex ecosystems by supporting distributed mood detection, aggregation of emotional patterns, and orchestration of generative responses.

\section{Use Cases}
\label{sec:usecases}


We illustrate the advantages of the proposed technology through a series of application scenarios, highlighting how emotion contagion orchestration operates in different interaction configurations and contributes to user experience and system-level outcomes.

\subsection{Use Case 1: Emotionally Aware Virtual Assistant for Customer Service}

\begin{figure}[h!]
    \centering
    \includegraphics[width=0.85\linewidth]{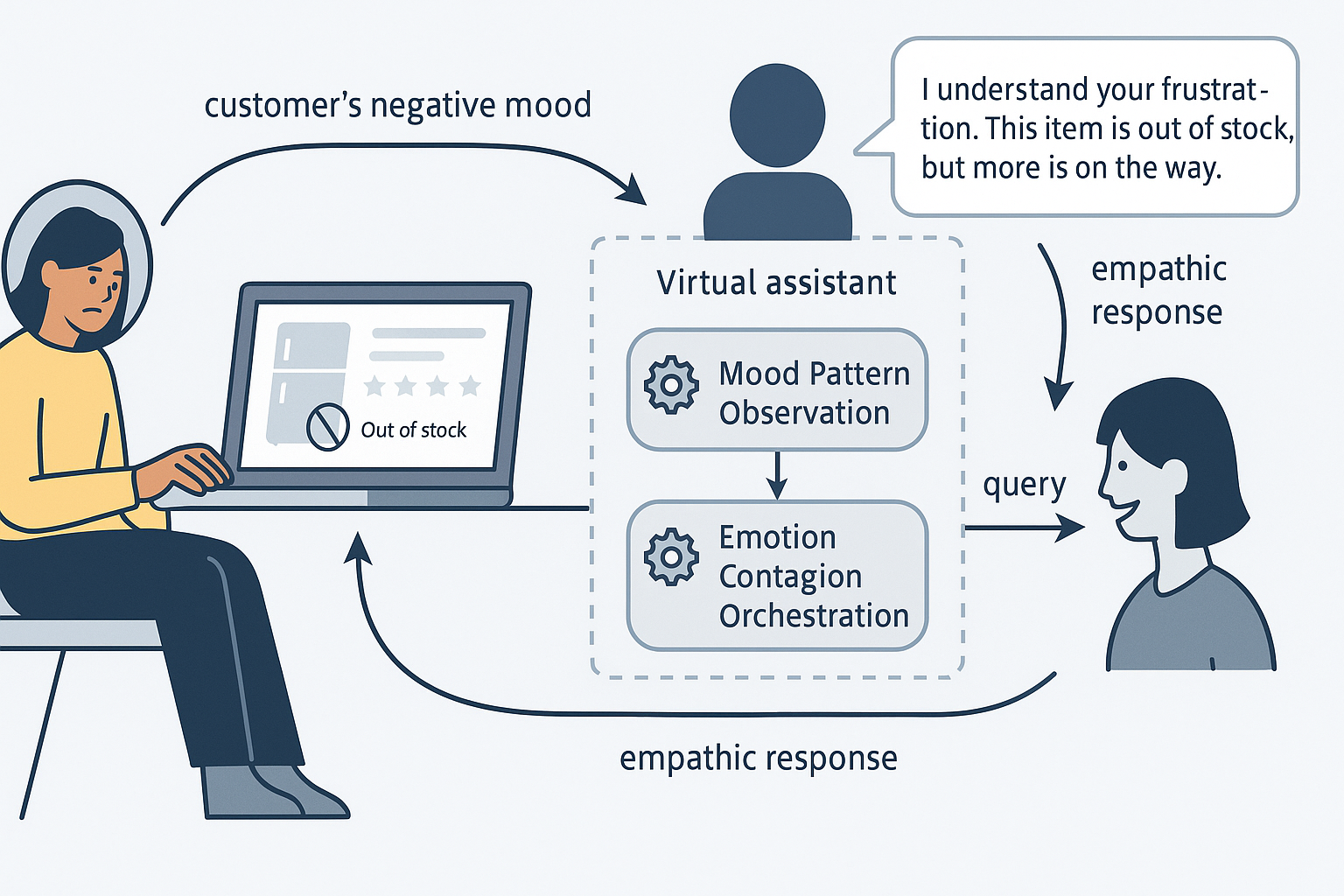}
    \caption{Emotionally Aware Virtual Assistant for Customer Service.}
    \label{fig:use-virtual-assistant}
\end{figure}

This scenario aligns with the configuration Human-Agent-Medium, where a user interacts directly with an agent through a shared communication medium. The agent has both emotional sensing and dialogue access, enabling emotion contagion.

A user engaged in product exploration encounters friction due to stock issues. The assistant detects early frustration through the Mood Pattern Observation Component and shares this with the broader agent network. The \textit{Mood Grouping Component} integrates trends from other users, and the orchestration layer guides the assistant to offer affectively supportive messages (e.g., encouragement, empathy, alternatives). Studies have shown that affective response modeling improves user satisfaction and reduces perceived friction in service interactions.

This orchestration elevates the quality of interaction beyond scripted customer support by introducing emotional intelligence. The result is improved user retention and brand trust, reflecting growing demand for emotionally responsive conversational agents.

\subsection{Use Case 2: Emotion-Aware Social Platform for Mental Health Support}

\begin{figure}[h!]
    \centering
    \includegraphics[width=0.85\linewidth]{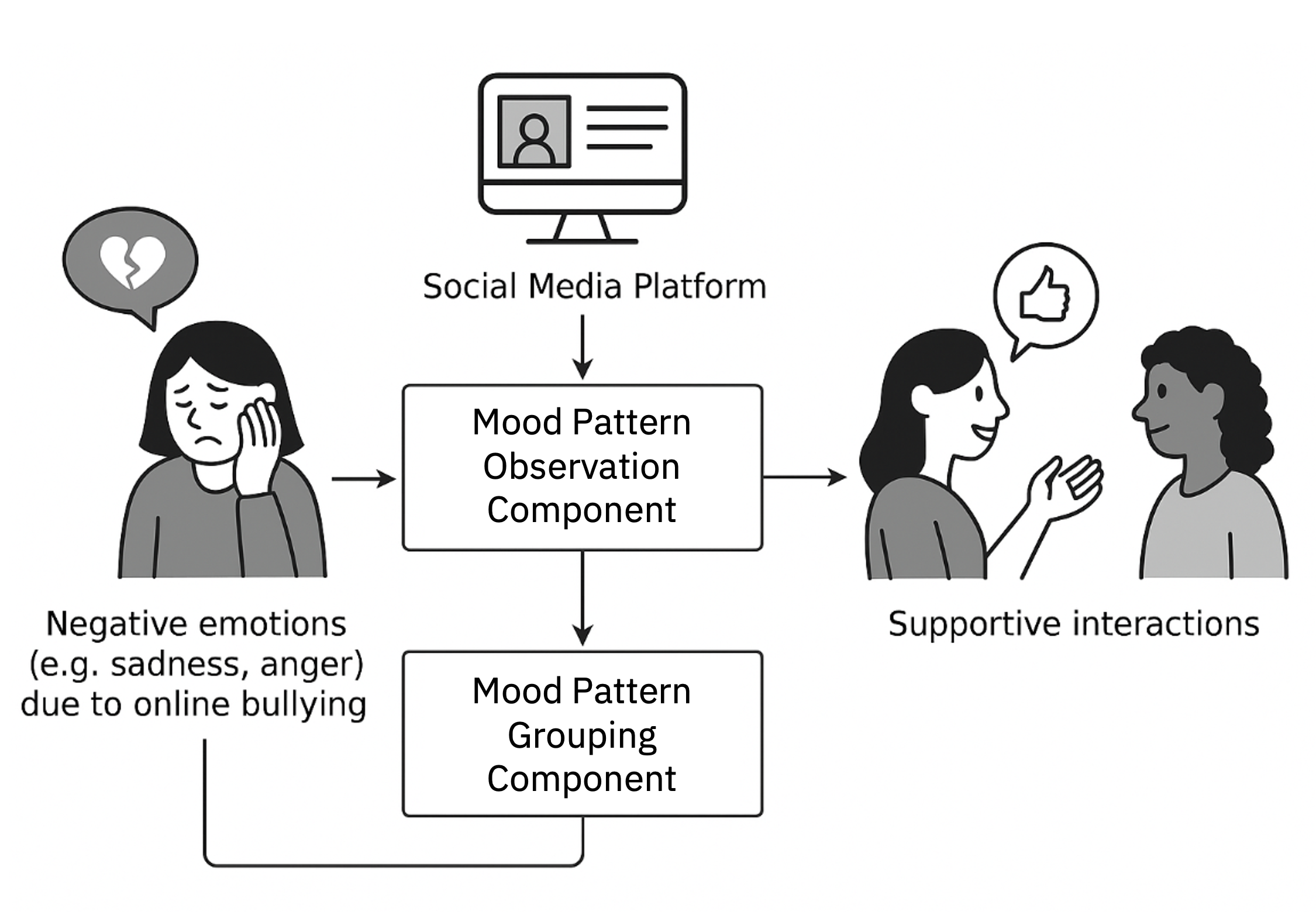}
    \caption{Emotion-Aware Social Platform for Mental Health Support.}
    \label{fig:use-mental-health}
\end{figure}

This use case maps to the configuration Agent-Human-Medium, where human users supported by a virtual assistant interact via a shared channel. These agents do not engage directly with the medium but assist individuals privately and may coordinate among themselves.

A user experiencing emotional distress due to online bullying exhibits signs of sadness or anger. The emotion-aware system detects these through indirect observation of posts and communication patterns. Agents supporting other users share emotional readings via the grouping component, which detects similar distress signals and emotional resilience pathways. The orchestration system facilitates interactions with empathetic peers and moderates distressing content visibility, fostering a safer digital environment.

This structure empowers agents to manage emotion contagion by identifying harmful spread patterns and guiding the user toward supportive engagements. The role of socially intelligent agents on mental wellness platforms is increasingly recognized as essential for emotional protection and collective well-being.

\subsection{Use Case 3: Emotionally Supportive Virtual Assistant in Healthcare Services}

\begin{figure}[h!]
    \centering
    \includegraphics[width=0.85\linewidth]{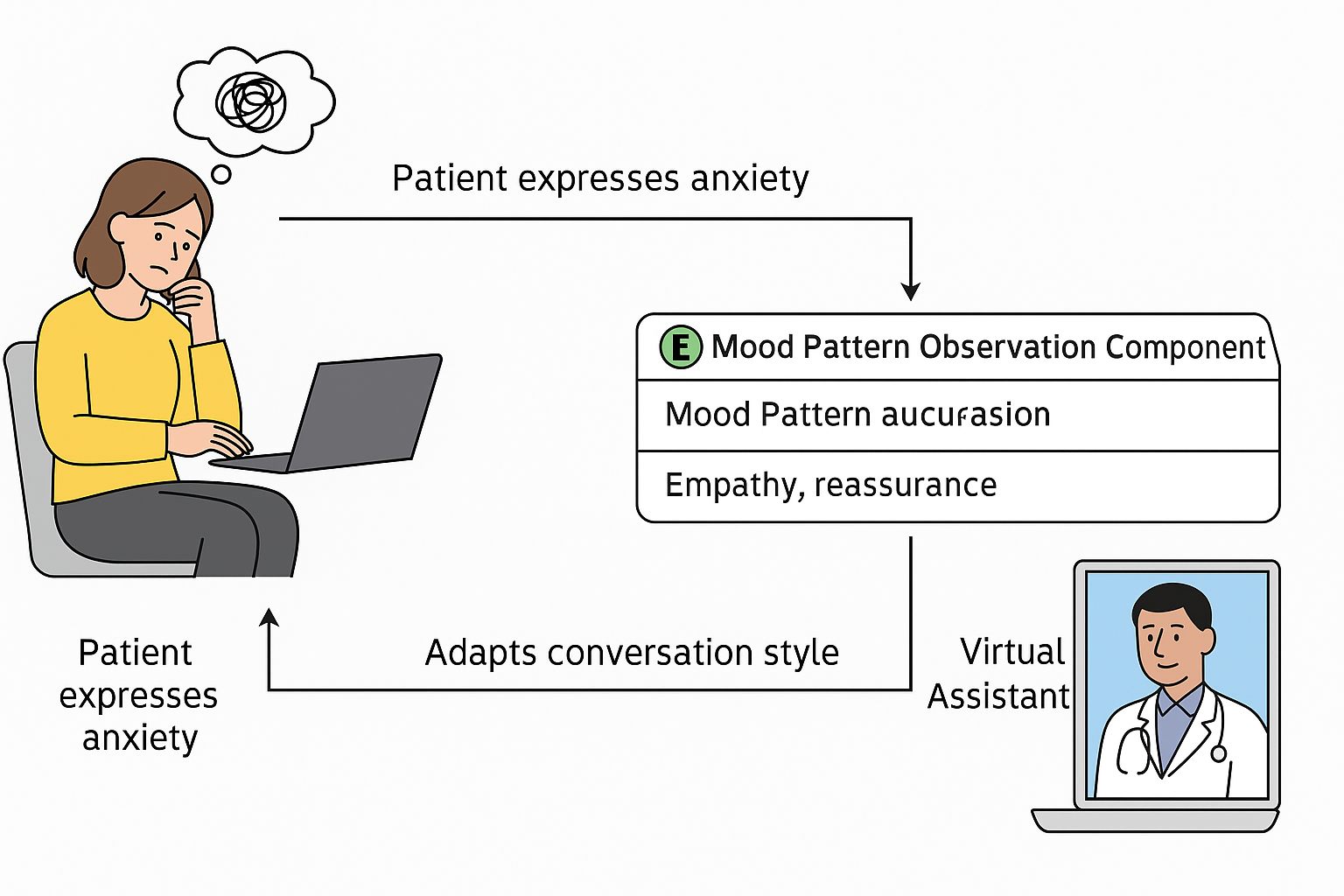}
    \caption{Emotionally Supportive Virtual Assistant in Healthcare Services.}
    \label{fig:use-healthcare}
\end{figure}

This scenario corresponds to configuration ``Humans and Agents Participating in Social Networks ([H|Ag]–M)'', in specific with the variation depicted in Fig.~\ref{fig:config-hma}(c). Here, agents have access to communication with both humans and the medium, and coordinate among themselves. This is a common structure in telemedicine platforms with multi-agent backends.

For instance, consider a scenario where a patient expresses anxiety during a teleconsultation. The assistant observes affective signals, shares mood data, and receives guidance to generate supportive messages and propose coping strategies. Coordination with other assistants reinforces consistent affective communication throughout care sessions. The system is able to suggest community interactions with patients showing similar affective patterns to build mutual support. Emotion-oriented dialogue generation has proven to be useful in building relationships and increasing patient trust in healthcare applications.

\subsection{Impact Across Use Scenarios}

The shared capability of collective emotional reasoning proves critical across these diverse scenarios. A system implementing an embodiment of our proposal will be able to coordinate affective alignment through  architecture-aware agent orchestration. The strategic involvement of agents across different configurations ensures flexibility and adaptiveness, key traits for real-world deployment This supports the formation of emotionally intelligent digital services that enhance satisfaction, increase trust, and reinforce the quality of experience for users across domains.

Our proposal relies on distributed emotional sensing and inference across a population of interacting agents, similar to how prior research has used agent-based sensing and statistical modeling to detect and recommend parking spots from smartphone data~\cite{koster2014recognition}. This architectural similarity underscores the broader applicability of decentralized sensing frameworks to environmental and affective coordination tasks in digital systems. Similarly, early work on grid-based intrusion detection~\cite{schulter2006towards} highlighted the importance of integrating localized anomaly detection with collective reasoning to manage distributed security threats, which is the inspiration for the approach our system adapts to the orchestration of emotional signals and collective agent response in multi-user contexts.

\section{Conclusions}
\label{sec:conclusions}


We introduced a framework that enables conversational agents to collaboratively sense, interpret, and influence emotional dynamics in multi-human, multi-agent environments. The proposed approach redefines affective computing by enabling a transition from isolated, reactive sentiment recognition to a collaborative and adaptive orchestration of group affect. Agents are designed to act as emotionally intelligent collaborators that contribute to a coordinated emotional landscape, rather than functioning as isolated conversational entities.

The framework introduces a novel architecture for scalable, real-time emotional alignment, through the integration of decentralized mood sensing, pattern grouping, interaction orchestration, and generative response generation. These applications reflect a growing need for systems that can respond to both individual users and the broader emotional trends influencing shared environments.

Our analysis of interaction configurations further clarifies where this technology offers the most leverage. The framework proves effective in situations where agents interact directly with users (H-M-Ag), operate as invisible personal assistants (Ag-H-M), or act as coordinated participants in communication environments. The system demonstrates its strongest impact in configurations where agents have both communication access and the ability to coordinate responses. In these setups, agents can share emotional context, collaboratively adjust responses, and dynamically steer the mood of the group in response to changing affective trends. This enables a level of collective emotional reasoning that traditional agent systems or isolated empathy modules cannot achieve.

This technology introduces new possibilities for designing emotionally effective and ethically responsible digital interactions. The framework lays the foundation for AI systems that prioritize emotional coordination, fostering psychological safety, resilience, and social trust in user interactions. As digital platforms continue to mediate social interaction, emotionally intelligent agent networks will play a central role in maintaining meaningful engagement and user well-being across application domains.


\section*{Acknowledgment}


This paper provides a scholarly description of Patent US20250094842\-A1 \cite{koch2025emotion}, which was conceived and authored while the authors were employed at International Business Machines, the current assignee of this patent application.

\bibliographystyle{IEEEtran} 
\bibliography{mybib}

\end{document}